\documentclass[pra,twocolumn,showpacs,superscriptaddress,aps]{revtex4}
\usepackage{amsfonts}
\usepackage{amsmath}
\usepackage{amssymb}
\usepackage{graphicx}
\usepackage{dcolumn}
\usepackage{times}
\usepackage{color}

\DeclareMathOperator{\sech}{sech}

\begin{document}

\title{Scattering of matter-waves in spatially inhomogeneous environments}

\author{F. Tsitoura}
\affiliation{Department of Physics, University of Athens,
Panepistimiopolis, Zografos, Athens 15784, Greece}
%%%
\author{P.  Kr\"{u}ger}
\affiliation{Midlands Ultracold Atom Research Centre, School of Physics \& Astronomy, 
The University of Nottingham, Nottingham, UK}
\author{P. G. Kevrekidis}
\affiliation{Department of Mathematics and Statistics, University of
Massachusetts, Amherst MA 01003-4515, USA}

\affiliation{Center for Nonlinear Studies and Theoretical Division, Los Alamos
National Laboratory, Los Alamos, NM 87544, USA}

\author{D. J. Frantzeskakis}
\affiliation{Department of Physics, University of Athens,
Panepistimiopolis, Zografos, Athens 15784, Greece}

\begin{abstract}

We study scattering of quasi one-dimensional matter-waves at an interface of 
two spatial domains, one with repulsive and one with attractive interatomic interactions. 
It is shown that the incidence of a Gaussian wavepacket from the repulsive to 
the attractive region gives rise to generation of a soliton train. 
More specifically, the number of emergent solitons can be controlled e.g.
by the variation of the amplitude or the width  of the
incoming wavepacket.
Furthermore, we study the reflectivity of a soliton incident from the attractive region 
to the repulsive one. We find the reflection coefficient numerically and employ analytical 
methods, that treat the soliton as a particle (for moderate and large amplitudes) or a 
quasi-linear wavepacket (for small amplitudes), to determine the critical soliton momentum 
-- as function of the soliton amplitude -- for which total reflection is observed. 

\end{abstract}


\maketitle

\section{Introduction}

For almost two decades, the study of nonlinear phenomena occurring in atomic Bose-Einstein condensates 
has experienced an enormous increase of interest \cite{book_fr,nl_car}. A prominent example, in the 
quasi one-dimensional (1D) setting, is the experimental observation of robust matter-wave solitons of 
the bright \cite{bs} and dark \cite{ds} type, and the study of their properties (see, e.g., the 
reviews \cite{abd} and \cite{djf} for bright and dark solitons, respectively). Such coherent 
nonlinear excitations of BECs are also interesting from the viewpoint of potential applications, 
ranging from coherent matter-wave optics to precision measurements and quantum information 
processing. Indeed, the formal similarities between nonlinear and matter-wave optics \cite{phil} 
indicate that coherent matter-waves may in principle be controlled similarly to their optical 
siblings in optical fibers, waveguides, photonic crystals, and so on \cite{kiag}.

In that respect, it is not surprising that there exist many works devoted to the 
manipulation of matter waves. Among various techniques that have been proposed, an experimentally 
tractable one refers to engineering the ``environment'' of the matter-wave, 
by magnetically \cite{mfr} or optically \cite{ofr} induced Feshbach resonances, 
which makes it possible to control the effective nonlinearity in the condensate. 
The application of such a ``Feshbach resonance management'' (FRM) technique \cite{ourFRM} 
in the temporal domain was used for the realization of matter-wave bright solitons  
by switching the interatomic interactions from repulsive to attractive \cite{bs}; it was also    
proposed as a means to stabilize attractive higher-dimensional BECs against collapse 
\cite{optnm1,optnm2} and to create robust quasi-1D matter-wave breathers \cite{ourFRM,optnm3}. 
On the other hand, the FRM technique in the spatial domain, which gives rise to 
the so-called ``collisionally inhomogeneous condensates'' \cite{g} with 
a spatially modulated nonlinearity, has also been extensively studied. In particular, 
novel phenomena and a variety of applications have been proposed in this context, including 
the adiabatic compression of matter-waves \cite{g,AS}, Bloch oscillations of matter-wave 
solitons \cite{g}, atomic soliton emission and atom lasers \cite{in4}, enhancement of transmittivity 
of matter waves through barriers \cite{in5}, formation of stable condensates exhibiting both 
attractive and repulsive interatomic interactions \cite{in8}, solitons in combined linear and 
nolinear potentials \cite{lnl}, generation of solitons \cite{21} and vortex rings \cite{22}, 
control of Faraday waves \cite{alex}, and many others.
A detailed recent review of such inhomogeneously nonlinear settings, especially
in the context of periodic (i.e., nonlinear lattice) variations
can be found in~\cite{karta}.

In this work, we study the scattering of matter-waves in a collisionally inhomogeneous environment. 
In particular, we consider a quasi-1D setting (whereby matter waves are oriented along 
the $x$-direction) and assume that the scattering length $a$ is piecewise constant for 
$x \ll 0$ and $x \gg 0$, taking respectively 
the values $-a_1 <0$ and $a_2>0$, and changes sign at $x=0$. In other words, we assume that 
the normalized scattering length $a(x)$ takes the form:
\begin{eqnarray}
a(x) &=& \frac{1}{2}\left[\left(\frac{a_2}{a_1}-1\right)+\left(\frac{a_2}{a_1}+1\right)
{\rm tanh}\left(\frac{x}{W}\right)\right], 
\label{a}
\end{eqnarray}
where $W$ is the spatial scale over which the transition from the asymptotic values 
$a_1$ and $a_2$ takes place. 
For the above setting, and in the framework of the mean-field approximation, 
we will investigate two different scattering processes; a description of 
our considerations and the organization of the paper are as follows. 

First, in Section II, we 
study the incidence of a nearly linear (Gaussian) wavepacket from the repulsive 
region ($x>0$) to the attractive region ($x<0$), and demonstrate the generation of 
a train of bright solitons. By numerically integrating the pertinent Gross-Pitaevskii (GP) 
equation, we determine the number of created solitons as functions of the initial data 
(amplitude, width and momentum of the incident wavepacket), 
as well as the difference of the values of the scattering length. 

In Section III, we study the reflectivity of a bright soliton from the scattering length 
interface; the soliton is assumed to exist and  travel from the 
attractive region ($x<0$) towards the repulsive region ($x>0$). 
We find numerically 
the reflection coefficient as a function of the soliton momentum and amplitude, and find 
that it has a step-like dependence on momentum for sufficiently weak solitons. 
In the case of total reflection, we use an analytical approximation (treating the 
soliton as a particle) and find the equation of motion for the soliton center. 
This equation is used to determine the critical value of momentum below which 
total reflection occurs, which turns out to depend linearly on the soliton amplitude. 
Additionally, for extremely weak solitons, employing results from linear quantum 
mechanics \cite{Sak}, we also find a (different) linear dependence of the critical 
momentum on the soliton amplitude. Both analytical estimates, for weak and strong solitons are 
found to be in very good agreement with the numerical results, with the
latter also encompassing a transition region between the two 
regimes. 

Finally, Section~IV concludes our findings and presents a number of
directions for future study.

\section{Reflectivity of a Gaussian wavepacket from the scattering length interface}

\subsection{Model and creation of a soliton train}
Our considerations start from the following Gross-Pitaevskii (GP) equation, which describes 
a quasi-1D BEC oriented along the $x$-axis \cite{book_fr,nl_car}:
\begin{eqnarray}
i\hbar\frac{\partial \Psi}{\partial t} &=& -\frac{\hbar^2}{2m}\frac{\partial^2 \Psi}{\partial x^2} 
+ 2\hbar \omega_\perp a |\Psi|^2 \Psi.
\label{gpe1}
\end{eqnarray}
Here $\Psi(x,t)$ is the mean-field order parameter, $m$ is the atomic mass, 
$\omega_\perp$ is the transverse confining frequency, and $a$ the s-wave scattering length 
[$a>0$ ($a<0$) corresponds to repulsive (attractive) interatomic interactions]. Considering a situation where  
proper spatially dependent 
%magnetic or optical 
fields close to Feshbach resonances are employed, we assume that 
the scattering length $a$ is piecewise constant for $x<0$ and $x>0$, taking the form 
\begin{equation}
a(x)=(1/2)[(a_2-a_1) + (a_1+a_2)\tanh(x/W)],
\end{equation}
where $W$ is the spatial scale over which the 
transition from the asymptotic value $-a_1<0$ (for $x/W \rightarrow -\infty$) 
to $a_2>0$ 
(for $x/W \rightarrow +\infty$) takes place. 

A strategy for developing a corresponding experimental implementation can be based on the interaction tunability of specific atomic species by applying external magnetic fields. For example, for cesium the s-wave scattering length $a$ changes sign through a zero-crossing at an external field strength of 17 G \cite{VuleticPRL1999}. Confining cesium atoms in an elongated trapping potential near the surface of an atom chip \cite{AtomChip} will allow for appropriate local engineering of $a$ to form steps of varying widths $W$, where the atom-surface separation sets a scale for achievable minimum step widths. The trapping potential can be formed optically, possibly also by a suitable combination of optical and magnetic fields, whereby care has to be taken as the magnetic field will influence both 
the external potential and the scattering length profile $a(x)$;
see e.g. the relevant discussion of~\cite{PRAHolmes2013}.

Normalizing time and space in Eq.~(\ref{gpe1}), as 
$t \rightarrow \omega_\perp t$ and $x \rightarrow x/ a_\perp$ 
(where $a_\perp =(\hbar/m \omega_\perp)^{1/2}$ 
is the transverse harmonic oscillator length), as well as the density as 
$|\psi|^2 \rightarrow 2\alpha_1|u|^2$, we cast Eq.~(\ref{gpe1}) 
into the following dimensionless form:
\begin{eqnarray}
i\frac{\partial u}{\partial t}+\frac{1}{2}\frac{\partial^2 u}{\partial x^2}-a(x)|u|^2 u &=& 0,
\label{psi}
\end{eqnarray}
where the function $a(x)$ is given by Eq.~(\ref{a}). 
For our analytical and numerical considerations 
below, we will use the values $a_2/a_1 =0.95$, and $W=0.01$ corresponding 
to an abrupt, step-like transition.

It is relevant to point out here that, generally, Eq.~(\ref{gpe1}) as well as its 
variants attempting to more adequately capture transverse degrees of freedom 
%(such as
(see the quasi-1D models of Refs.~\cite{NPSE} and~\cite{delgado}),
suggest that the variations/modulations 
in transverse trapping strength can be used
in a way equivalent to longitudinal variations of the scattering length.
This idea has been used even in a quantitative fashion, e.g., to explain
the phenomenology of the formation of Faraday wave patterns -- cf. Ref.~\cite{nicolin} 
(and relevant work in Ref.~\cite{alex}). Nevertheless, this type of consideration is
not applicable in the present setting, given the sign changing
nature of the nonlinearity.

We now assume that a Gaussian wavepacket of amplitude $U_0$ and width $l$, initially 
located at $x=x_0>0$ (i.e., in the repulsive region), moves towards the attractive region. 
The specific form of the wavepacket, which is used as an initial condition 
for Eq.~(\ref{psi}) in our simulations, is:
\begin{eqnarray}
u(x,0) &=& U_0\exp\left(-\frac{\left(x-x_0\right)^2}{l^2}\right)\exp\left({-iKx}\right),
\label{gaussian}
\end{eqnarray}
where $K$ is the initial momentum of the wavepacket. 
Notice that this form of the wavepacket approximates the ground state profile 
in the case of relatively small atom numbers (corresponding to a weak nonlinearity).

\begin{figure}[tbp]
\centering
\includegraphics[scale=0.37]{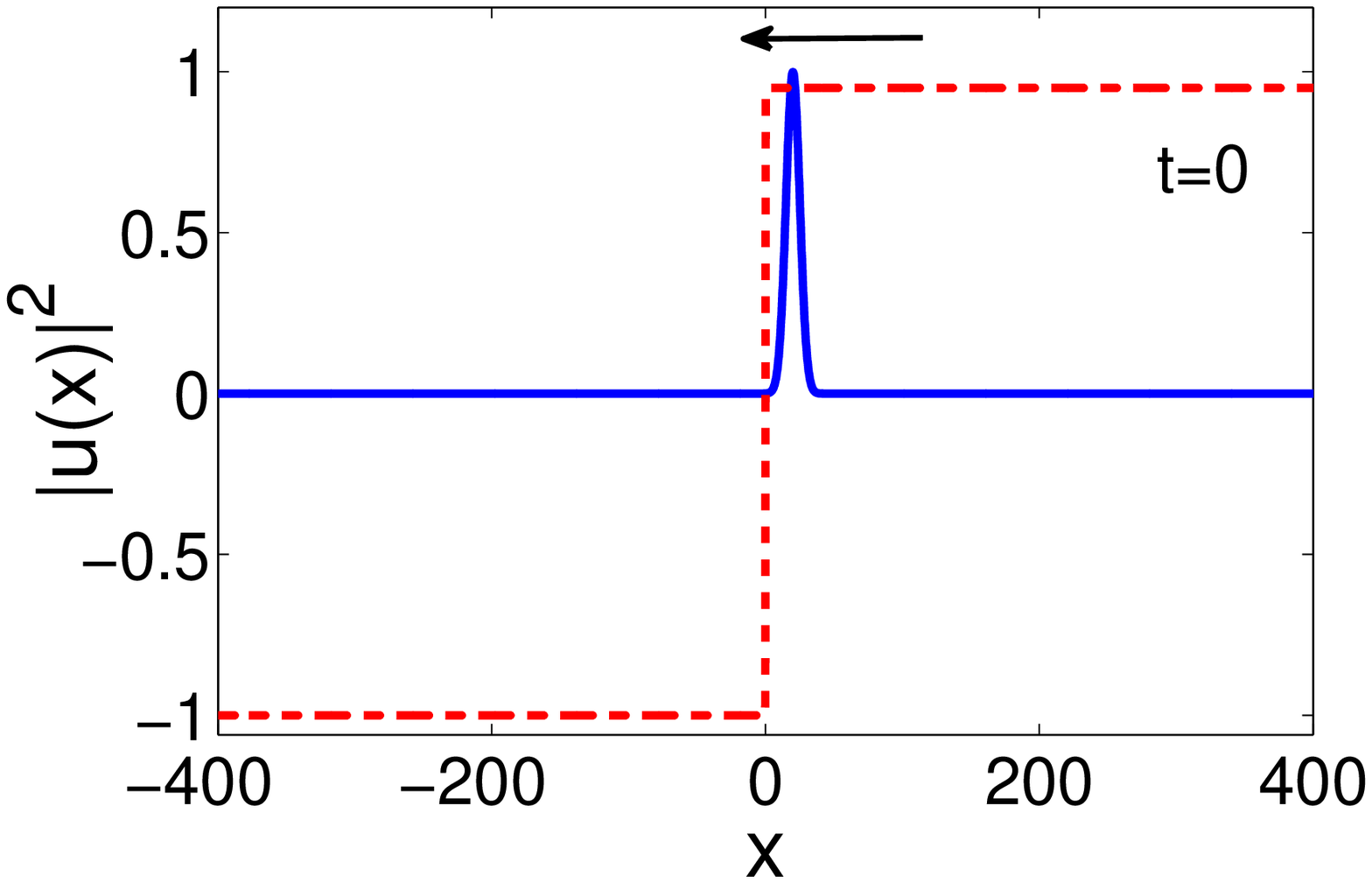}
\includegraphics[scale=0.37]{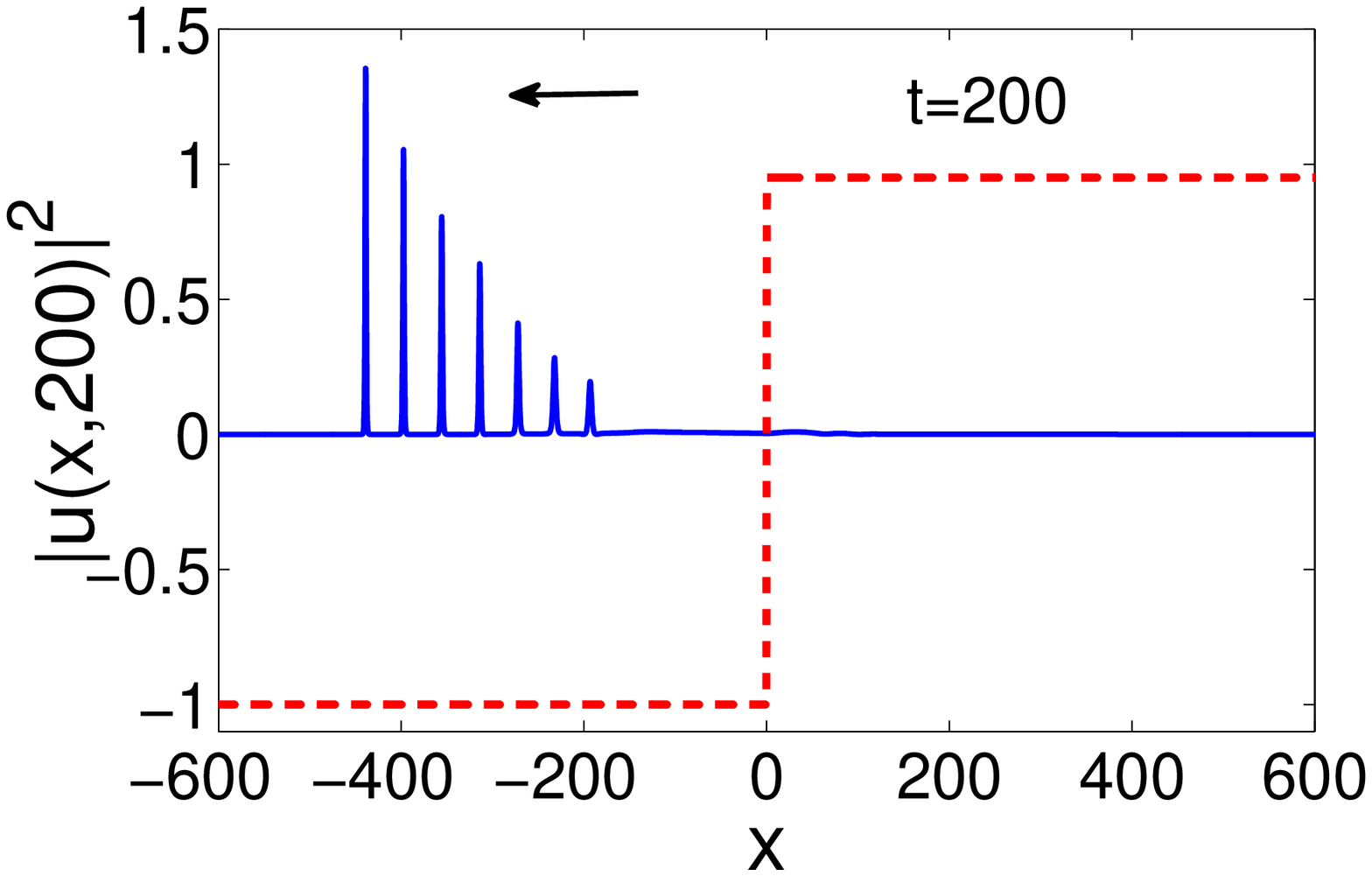}
\includegraphics[scale=0.37]{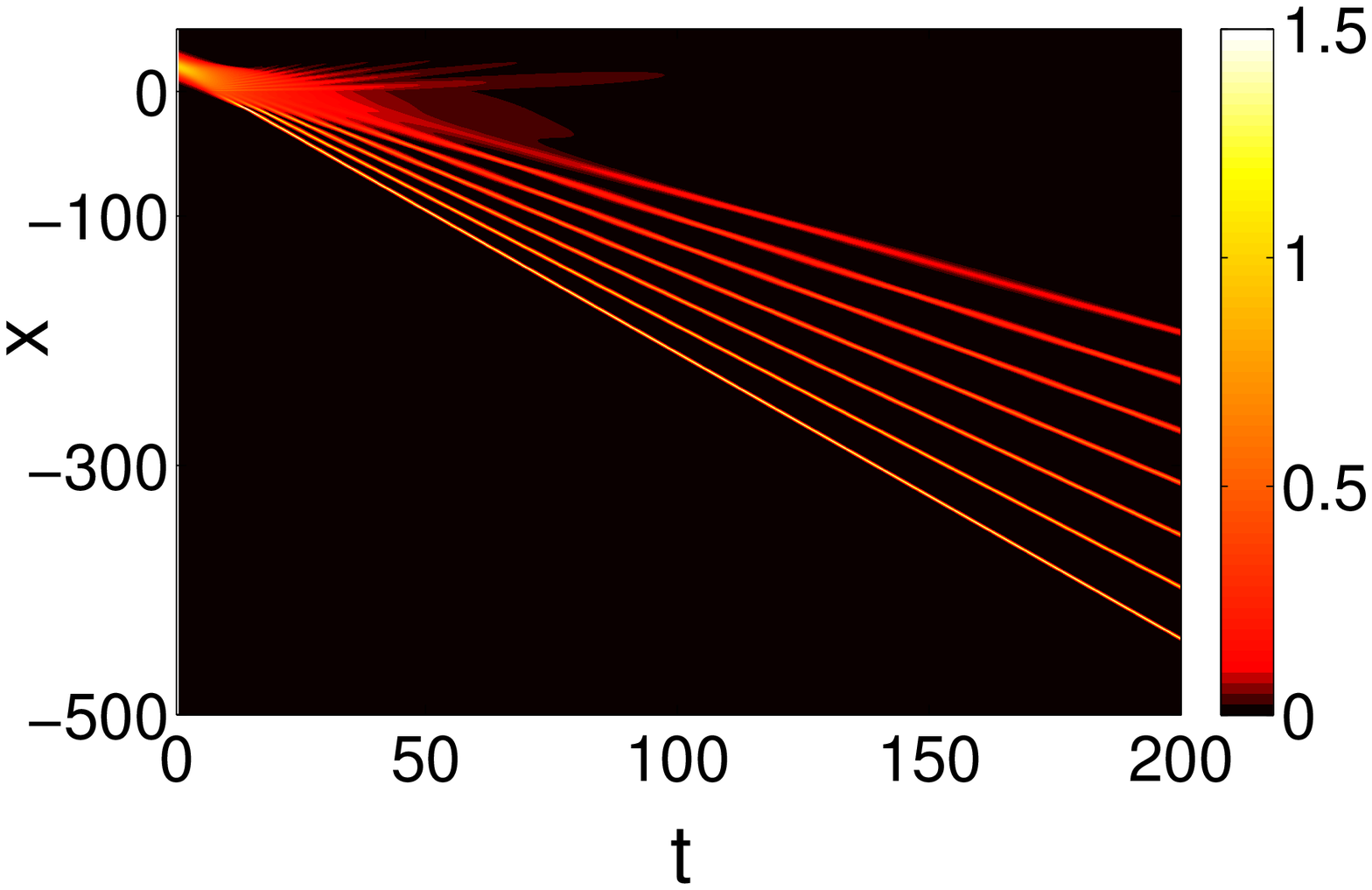}
\caption{(Color online) 
Motion of a Gaussian wavepacket from the region with repulsive interactions to the region with 
attractive interactions. Top and middle panels show, respectively, the density profiles of the 
wavefunction at at $t=0$ and $t=200$ [solid (blue) lines]; the normalized scattering length profile 
for $W=0.01$ is also depicted [dashed (red) line]. The bottom panel 
represents a contour plot showing the 
evolution of the density; middle and bottom panels clearly show the creation of a bright soliton 
train. Parameter values are: $U_0=1$, $x_0=20$, $l=10$, $K=1.5$, and $a_2/a_1 =0.95$.
}
\label{Fig5}
\end{figure}

Using the parameter values $x_0= 20$, $l=10$, and $K=1.5$ 
(as well as $U_0=1$, $W=0.01$ and $a_2/a_1 =0.95$), 
we depict the corresponding configuration in the top panel of Fig.~\ref{Fig5}. In  
the middle and bottom panels of the same figure, we show the subsequent dynamics: 
it is observed that the wavepacket is transmitted through the discontinuity of the 
scattering length at $x\approx 0$ and, after entering the region with attractive interactions, it 
transforms into a train of bright solitons. Notice that the soliton generation process is such that 
each generated soliton is larger than the one that will be generated at a later time. This is due 
to the fact that once a portion of the condensate enters the attractive side 
and is self-organized 
into a soliton, the number of atoms of the wavepacket on the repulsive side is decreased and, thus, 
a smaller soliton will be generated next. It is interesting to observe that the ratio of the 
velocities of two adjacent solitons in the train is constant; as a result, for each certain 
time instance, the distance between adjacent solitons is the same. 
Note that the results on the generation -- and characteristics -- of the atomic soliton train described above 
are reminiscent to the ones found in Ref.~\cite{in4}, but by means of a 
somewhat different physical mechanism: in that work, 
the soliton train was produced via a sufficiently deep spatially-dependent nonlinearity which 
acted on a trapped Gaussian wavepacket (existing between a region of 
vanishing and that of negative scattering length). The depth of the
(abrupt) negative step was found to control the number of emitted solitary
waves.

\begin{figure}[tbp]
\centering
\includegraphics[scale=0.4]{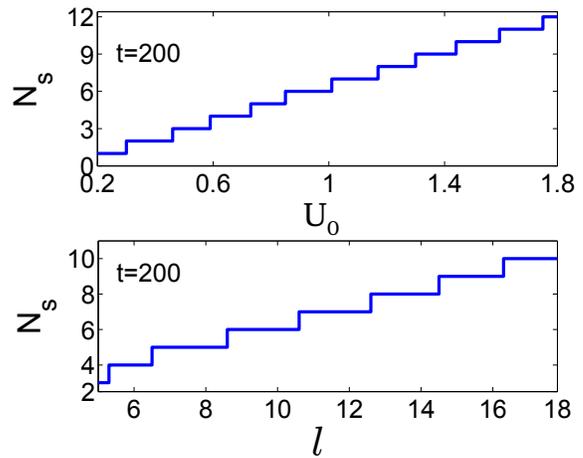}
\caption{(Color online) 
The top and bottom panels show, respectively, the number of solitons, 
$\rm N_s$, observed at $t=200$, as a function of the initial amplitude $U_0$ (for fixed $l=10$) 
or as a function of the initial wavepacket width $l$ 
(for fixed $U_0=1$) of the Gaussian wavepacket. The other parameter values are 
$W=0.01$, $a_1=1$, $a_2=0.95$, $K=1.5$,  and $x_0=20$.}
%The left and right panels show, respectively, the number of solitons, 
%$\rm N_s$, observed at $t=200$, as a function of the initial amplitude (for fixed $k=1.5$) 
%or as a function of the initial wavepacket width 
%(for fixed $a_1=1$ and $a_2=0.95$) of the Gaussian wavepacket. The other parameter values are 
%$W=0.01$, $a_1=1$, $a_2=0.95$, $l=10$,  and $x_0=20$.}
\label{plot1}
\end{figure}

We find that the number of the created solitons, $N_s$, depends on the momentum $K$, 
the amplitude $U_0$ and the width $l$ of the Gaussian 
wavepacket, as well as the height of the interface $a_2/a_1$. 
Results pertaining to the count of the soliton number are shown in Fig.~\ref{plot1}: 
larger initial amplitudes and/or widths of the wavepacket 
result in a larger number of solitons. On the other hand, increasing the initial momentum $k$ 
and/or the height $a_2/a_1$ of the interface the number of solitons seems
to have a weaker effect on the process; for the particular example 
shown in Fig.~\ref{Fig5}, the number of 
%created is constant and 
solitons is $7$ at time $t=200$. Here, we should note that for the counting of the number 
of solitons, we have included only solitons of amplitudes at least $10\%$ of the first created soliton.

\section{Reflectivity of a soliton from the scattering length interface}

Next, we consider the reflectivity of a bright soliton at the scattering length interface. 
We assume, in particular, that a bright soliton moves from the attractive 
($x<0$) to the 
repulsive region ($x>0$) and is, thus, scattered at the interface, at $x=0$, caused by the change of the 
sign of the nonlinearity. This dynamical scenario is, effectively, complementary to the one studied 
in Section~II. 
%[recall that in the latter we had studied the incidence of a localized wavepacket 
%from the repulsive  ($x>0$) to the attractive region ($x<0$)].

%
\begin{figure}[tbp]
\centering
\includegraphics[scale=0.4]{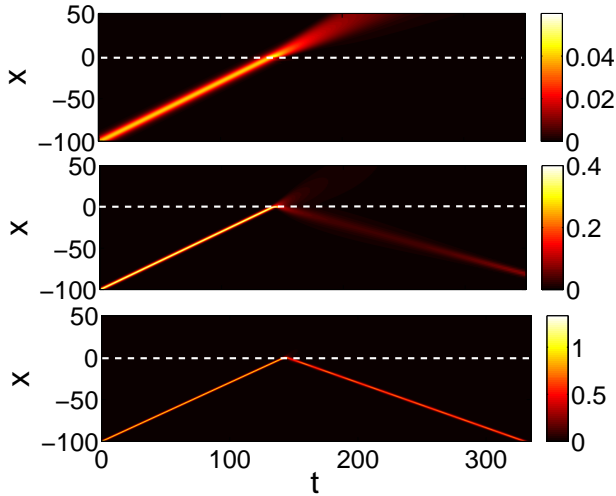}           
\caption{(Color online) Contour plots showing the evolution of the density of a 
bright soliton of initial velocity $k=0.7$, scattered at the interface between 
the attractive and repulsive region, at $x=0$ [depicted by the dashed (white) line]. 
From top to bottom, the amplitude of the soliton is $\eta=0.2$, $\eta=0.6$ and $\eta=1$. 
Top (bottom) panel corresponds to total 
transmission (reflection); middle panel shows partial reflection. 
}
\label{cont_R}
\end{figure}

The bright soliton propagating in the attractive region has the form:
\begin{eqnarray}
u(x,t) &=& \eta {\rm sech} \left[\eta\left(x-x_0(t)\right)\right]
\exp\left(i\left(kx-\omega t\right)\right),
\label{u1}
\end{eqnarray}
where $\eta$, $k$, $x_{0}$ and $\omega$ respectively denote the amplitude, velocity, 
initial position and frequency of the soliton. 
Then, we numerically integrate Eq.~(\ref{gpe1}) with the 
initial condition taken as
%from Eq.~({\ref{u}}) %at time $t=0$, 
\begin{eqnarray}
u(x,0) &=& \eta {\rm sech} \left[\eta\left(x-x_{0}(0)\right)\right]\exp\left(ikx\right)\exp\left(i\phi\right),
\label{u}
\end{eqnarray}
and observe the dynamics of the scattering process. Typical outcomes 
are shown in Fig.~\ref{cont_R}; in all cases, we fix the initial soliton momentum, at $k=0.7$, 
and vary the amplitude $\eta$. We observe that if the soliton amplitude is 
sufficiently small (large) then total transmission (reflection) is found --cf. top (bottom) panel 
of the figure for $\eta=0.2$ ($\eta=1$). On the other hand, for a moderate value of $\eta$ 
(e.g., $\eta=0.6$ --cf. middle panel) the soliton is partially transmitted and reflected.

The soliton reflectivity can be calculated numerically 
upon determining the reflection coefficient $R$, defined as the number of atoms 
remaining in the $x<0$ (attractive) region over the number of atoms of the incident 
soliton. Taking into regard that the latter is given by
$\int_{-\infty}^{\infty}|u(x,0)|^2 dx = 2\eta$, we can express $R$ as:
\begin{eqnarray}
R=\frac{1}{2\eta}\int_{-\infty}^{0}|u(x,t_\star)|^2 dx.
\label{Rnum}
\end{eqnarray}
Here, $t_\star$ is a time sufficiently large such that the reflected and transmitted parts 
of the soliton are spatially well separated; this separation is set by a spatial region of extent 
$\Delta x \approx k t_\star$ around $x=0$ and, accordingly, $t_\star$ is appropriately chosen for 
each individual numerical experiment. 

\begin{figure}[tbp]
\centering
\includegraphics[scale=0.4]{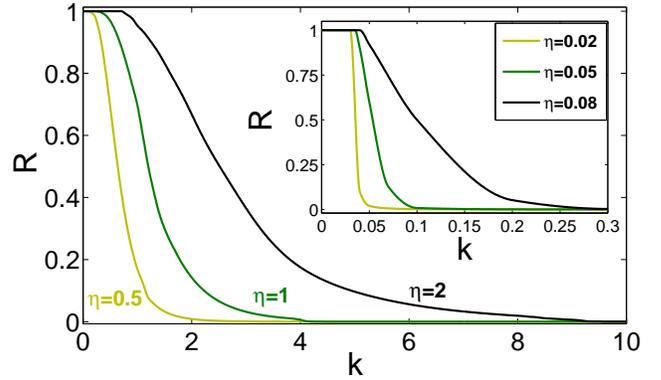}
\caption{(Color online) The reflection coefficient 
$R$ as a function of the initial soliton momentum 
$k$, for various values of large soliton amplitudes $\eta$. 
The inset shows cases corresponding to weak solitons.}
\label{Rk}
\end{figure}

Figure~\ref{Rk} shows the reflection coefficient as a function of the initial soliton momentum  
(for $0 \leq k \leq 10$) and various values of the soliton amplitude $\eta$. 
We observe that when $\eta$ is increased, the respective reflection coefficient curves 
drift towards larger momentum values and the curves become smoother: the transition from total 
reflection to total transmission becomes less sharp. This means that the interval of momenta for 
which partial transmission and reflection occur (as in the middle panel of Fig.~\ref{cont_R}) 
increases with increasing soliton amplitude. 

From the above discussion, it is obvious that the soliton keeps its particle-like character 
only in the case where it is totally reflected (cf. bottom panel of Fig.~\ref{cont_R}): 
in the cases of total or partial transmission, the soliton is not supported in the repulsive 
regime and it is eventually destroyed. We can thus adapt the 
particle picture for the soliton dynamics in the total reflection regime, and describe analytically 
the soliton trajectory and its reflectivity properties. Our approach
based on the center of mass (defined below) extends the corresponding
considerations of Ref.~\cite{nogami}, where a similar methodology was developed 
for the case of a linear step potential.

\begin{figure}[tbp]
\centering
\includegraphics[scale=0.3]{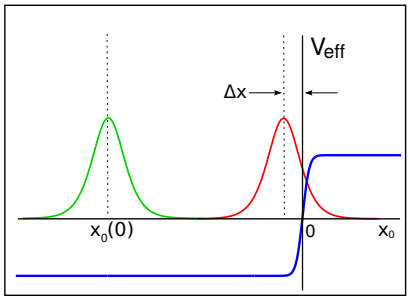}
\caption{(Color online) A sketch of the effective potential $V_{\rm{eff}}$ (blue line) 
as a function of $x_0$. Shown also is the soliton, initially located at $x_0(0)$ [left (green) curve], 
far from the scattering length interface, and in close proximity to the interface [right (red) curve], 
where the location of its center is $x_0(|\Delta x|)$. 
}
\label{hwhm2}
\end{figure}

We start with the soliton's center of mass, given by: 
\begin{eqnarray}
{\overline{x}}=\int_{-\infty}^{\infty}x|u(x,t)|^2 dx,
\label{cm}
\end{eqnarray}
which is connected with the soliton momentum 
$P=\int_{-\infty}^{\infty} \left(u u_x^{*}-u^{*}u_x\right) dx$ 
through the equation ${d\overline{x}}/dt=P$. Then, differentiating the latter expression 
with respect to $t$, and using Eq.~(\ref{psi}), it is straightforward to derive the following 
equation of motion for $\overline{x}$: 
\begin{eqnarray}
\frac{d^2 {\overline{x}}}{dt^2} 
&=& \frac{1}{2} \int_{-\infty}^{\infty} \frac{da(x)}{dx} |u(x,t)|^4 dx.
\label{xcm1}
\end{eqnarray}
The integral on the right-hand side of Eq.~(\ref{xcm1}) can be calculated in an analytical 
form, upon approximating $a(x)$ [cf. Eq.~(\ref{a})] by a Heavyside function in the limiting case 
where $W \ll 1/\eta$. Then, $da(x)/dx$ is approximated by a delta function, and integrating the 
right-hand side of Eq.~(\ref{xcm1}), we end up with the following result: 
\begin{eqnarray}
\frac{d^2 {\overline{x}}}{dt^2} &\approx & 
-\frac{1}{2}\left(1+\frac{a_2}{a_1}\right)\eta^4 {\sech}^4 \left(\eta {x_0}\left(t\right)\right).
\label{xcm2}
\end{eqnarray}
Then, taking into regard that the soliton center is connected with the center of mass through the 
equation $\overline{x}=2\eta{x_0}$, we can express Eq.~(\ref{xcm2}) as follows: 
\begin{eqnarray}
\frac{d^2 {x_0}}{dt^2} &=& -\frac{dV_{\rm{eff}}}{d{x_0}},
\label{eqmot}
\end{eqnarray}
where the effective potential $V_{\rm{eff}}$ is given by:
\begin{eqnarray}
V_{\rm{eff}}({x_0}) &=& \frac{1}{12} \left(1+\frac{a_2}{a_1}\right) \eta^2 \nonumber \\
&\times& \left[3 {\tanh}\left(\eta x_0 \right)-{\tanh}^3\left(\eta x_0 \right) \right].
\label{V2}
\end{eqnarray}
Equation (\ref{eqmot}) shows that the soliton can be regarded as a Newtonian 
unit-mass particle, which evolves in the presence of the effective potential $V_{\rm eff}$; 
the latter, has a shape of a step-like barrier, as depicted in Fig.~\ref{hwhm2}. 
Thus, according to this particle picture, 
the soliton will be totally reflected if its initial energy $E_{\rm s}$ is 
less than the ``height'' of the barrier. 
Since the soliton is expected to interact with the effective potential only through 
its exponential leading tail, the soliton center is anticipated to never reach the interface at 
$x=0$, but it will approach it only up 
to a distance roughly equal to the half-width at half-maximum (HWHM) of the soliton; 
the above situation is schematically illustrated in Fig.~\ref{hwhm2}.
Thus, taking into regard that the soliton's HWHM, denoted by $\Delta x$, is connected with the 
inverse width $\eta$ through the equation $\Delta x =\ln (1+\sqrt{2})/\eta$, we can find that the 
relevant barrier height is given by $V_{\rm eff}(\Delta x)-V_{\rm eff}(x_0(0))$, 
where $x_0(0)$ is the initial soliton position. 
According to the above arguments, the soliton will be totally reflected if the initial soliton energy 
is less than (or equal to) the effective barrier height, namely:
\begin{equation}
E_{\rm s} \equiv \frac{1}{2}k^2 + V_{\rm eff}(x_0(0)) \le V_{\rm eff}(\Delta x)
%-V_{\rm eff}(x_0(0)).
\label{solen}
\end{equation}

\begin{figure}[tbp]
\centering
\includegraphics[scale=0.4]{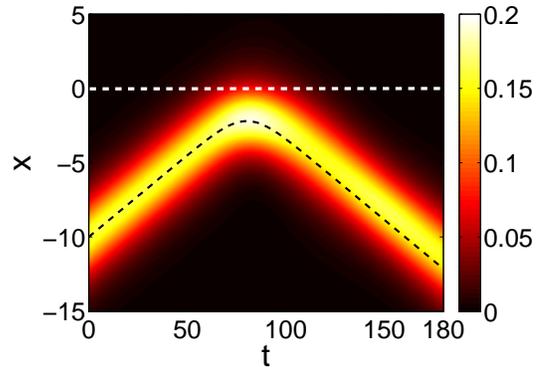} 
\caption{(Color online) Contour plot showing 
the space-time evolution of the motion of the soliton center. 
The bright soliton is initially placed at $x_0(0) = -10$, while its amplitude 
and momentum are $\eta=0.4$ and $k=0.1$. 
The dashed (black) curve represents the analytical result of Eq.~(\ref{eqmot}), 
while the horizontal (white) line depicts $x=0$.}
\label{traj}
\end{figure}

We have numerically checked the validity of this analysis by 
comparing, at first, the numerically obtained soliton 
trajectory [by means of direct numerical integration of Eq.~(\ref{psi}) in the case of total 
reflection] with the approximate analytical 
result of Eq.~(\ref{eqmot}). A typical example, corresponding to a soliton amplitude 
$\eta=0.4$ and momentum $k=0.1$, is shown in Fig.~\ref{traj}. There, the 
numerical result is displayed in the form of a contour plot for the evolution of the soliton density, 
as well as the analytical result of Eq.~(\ref{eqmot}) -- cf. dashed line in the figure.
Note that similar results were obtained for soliton amplitudes $0.2<\eta<2$. 
It can be seen that the dashed line follows with a fairly good accuracy the evolution of the 
soliton center.  The slight discrepancy observed can be 
explained as follows: the tail of the bright soliton, in case of total 
reflection (cf. Figs.~\ref{hwhm2} and \ref{traj}), interacts with the interface 
and enters the repulsive area, and eventually comes back to the 
attractive region. 
This effect, which cannot be explained via the particle approach, 
causes a slight shift in the soliton 
trajectory. Thus, the trajectory obtained from 
Eq.~(\ref{xcm1}) has naturally a slight discrepancy 
for any soliton amplitude $\eta$. 

\begin{figure}[tbp]
\centering
\includegraphics[scale=0.4]{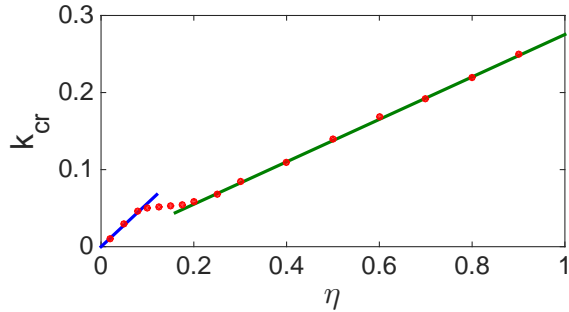}
\caption{(Color online) The critical momentum $k_{\rm{cr}}$, below which total 
reflection occurs, as function of the soliton amplitude $\eta$. The (red) dots 
depict the numerical result, while the left (blue) and right (green) straight lines 
correspond to the analytical predictions of Eqs.~(\ref{klinear}) and (\ref{knonlin}), 
respectively. The parameter values are $a_1=1$, $a_2=0.95$ and $W=0.01$.}
\label{etak}
\end{figure}

Next, employing Eq.~(\ref{solen}), it is possible to derive analytically the critical value 
of the initial momentum $k_{\rm cr}$ [when the equality in Eq.~(\ref{solen}) holds], 
for which total reflection occurs, as a function of the 
soliton amplitude and the parameters characterizing the scattering length profile.
The result is:
\begin{eqnarray}
k_{\rm{cr}} &=& \left[ \frac{1}{3}\left(1+\frac{a_2}{a_1}\right)(1+C)\right]^{1/2}~\eta,
\label{knonlin}
\end{eqnarray}
where constant $C=(1/2)[~3 \tanh(\eta \Delta x )-
\tanh^3 (\eta \Delta x )]$ and $\eta \Delta x = -\ln (1+\sqrt{2}) \approx -0.88$. 
Note that Eq.~({\ref{knonlin}}) suggests a linear dependence of $k_{\rm{cr}}$ on $\eta$, 
which is confirmed by our numerical simulations. Indeed, as shown in Fig.~\ref{etak}, for 
solitons of sufficiently large amplitudes, i.e., for $\eta \gtrsim 0.2$, this 
analytical prediction [depicted by the solid (green) straight line] is in an excellent 
agreement with the numerical result for $k_{\rm{cr}}$ [depicted by the (red) dots]. Notice 
that the numerically obtained values for $k_{\rm{cr}}$ are calculated so that the 
respective reflection coefficient values become less than unity by a factor of $10^{-3}$; however, we note here that the results presented are only 
weakly sensitive to the selection  of the particular threshold.

For weaker solitons it is expected that our analytical approximations described 
above should be less accurate: this is due to the fact that for small values of $\eta$, the nonlinearity 
becomes extremely weak, and thus a linear description of the problem would be more appropriate. 
In such a case, the soliton can be treated as a linear wavepacket, which is scattered from an 
effective step barrier; the latter, is basically formed by the step-like 
change of the scattering length profile. Then, the reflection coefficient can be approximated 
from the corresponding linear problem \cite{Sak} as follows:
\begin{eqnarray}
R=1-\frac{4\sqrt{(E-V_0)E}}{(\sqrt{E}+\sqrt{E-V_0})^2}, 
\label{th}
\end{eqnarray}
where $E$ and $V_0$ denote, respectively, the energy of the wavepacket and 
the height of the effective potential barrier. Notice that Eq.~(\ref{th}) stands for 
plane waves; however, it can still provide a reasonable approximation 
as long as the soliton width $\eta^{-1}$ is sufficiently large, i.e., for sufficiently 
weak solitons. In our case, the soliton energy is given by 
(see, e.g., Ref.~\cite{nl_car}):
\begin{eqnarray}
E = \eta k^2 -\frac{1}{3}\eta^3,
\label{E}
\end{eqnarray}
while the strength of the effective barrier potential is given by:
\begin{eqnarray}
V_0 &=& \frac{1}{2}\int_{- \infty}^{\infty} a(x) |u|^4 dx
=\frac{1}{3} \left(\frac{a_2}{a_1}-1\right)\eta^3.
\label{V0}
\end{eqnarray}
Then, total reflection, i.e., $R=1$ in Eq.~(\ref{th}), occurs for $E=V_{0}$; the latter 
equation leads to the following result for the critical momentum $k_{\rm{cr}}$:
\begin{eqnarray}
k_{\rm{cr}} &=& \left(\frac{a_2}{3a_1}\right)^{1/2} \eta.
\label{klinear}
\end{eqnarray}
The above approximate analytical result, which is relevant to weak solitons, 
also shows a linear dependence of $k_{\rm{cr}}$ on $\eta$ and is in a very good agreement 
with the numerical results, as shown in Fig.~\ref{etak} for $\eta \lesssim 0.1$.

In summary, we capture the regime of small
$\eta$ by means of the linear/wave theory, and the 
regime of large $\eta$ by
our soliton particle theory, while between the two we interpolate via the use
of numerical computations as shown in Fig.~\ref{etak}.

\section{Conclusions}

In this work, we studied the scattering of quasi-1D matter-waves in a spatially inhomogeneous 
environment, characterized by a piecewise constant profile of the scattering length $a$, 
such that $a=-a_1<0$ for $x<0$, $a=a_2>0$ for $x>0$, and $a$ changes sign at $x=0$. 
This way, in the region $x<0$ ($x>0$) the interatomic interactions are attractive (repulsive).
We investigated two different dynamical scenarios: 
\begin{itemize}
\item[(i)] the scattering of a quasi-linear (Gaussian) wavepacket at the scattering length 
interface, with the wavepacket traveling from the repulsive to the attractive region, and 
\item[(ii)] the scattering of a matter-wave bright soliton at the scattering length 
interface, with the soliton traveling from the attractive to the repulsive region. 
\end{itemize}

In case (i), we found that when the wavepacket enters the attractive region it evolves 
into a train of bright solitons. The soliton train is such that each generated soliton is 
larger than the one that will be generated at later times, 
while the distance between adjacent solitons is the same. 
We counted the number of the created solitons, as a function of the wavepacket's initial 
characteristics (momentum, amplitude, and width) and the height of the nonlinearity 
interface $a_2/a_1$, and found that larger initial amplitudes and/or widths of the 
initial wavepacket result in a larger number of solitons.

For case (ii), we found that the incidence of the soliton at the 
scattering length interface generally leads to total transmission, total reflection, or 
partial transmission/reflection. 
The reflection coefficient was determined numerically as a function of the 
initial soliton momentum, for different soliton amplitudes.  
For sufficiently weak solitons, we found an almost abrupt change 
from total transmission to total reflection, effectively associated
with the linear phenomenology in a step potential.
For stronger solitons, the reflection 
coefficient featured a smoother dependence on the momentum. 
%Numerically, we find a smooth crossover between these two regimes, which can 
%be interpreted as a gradual continuous change of the soliton from being 
%dominated by wave-like to particle-like properties.

We also developed analytical approximations -- that treated 
the solitons as particles (for large amplitudes) or linear wavepackets (for small amplitudes) 
-- to determine the critical value of soliton momentum, $k_{\rm cr}$, 
below which total reflection occurs. We found that $k_{\rm cr}$ depends linearly on 
the soliton amplitude, but with different slopes in the purely nonlinear and 
the quasi-linear regimes. Numerically, we find a smooth crossover between these two regimes, which can 
be interpreted as a gradual continuous change of the soliton from being 
dominated by wave-like to particle-like properties.
Our analytical predictions were found to be in very good
agreement with the corresponding numerical results. 

There are numerous directions that may be worth considering for future
efforts. One of these is to consider the possibility of multiple steps
and their interplay. Another is to examine the interplay of the nonlinear
step with an external linear potential or with a non-trivial background 
(e.g. on the repulsively interacting side which can support such
a background) and to explore the dynamics of incident wavepackets
in such settings. Potentially, probing the soliton dynamics in
such configurations could be utilized towards retrieving quantitative
information about the nature of linear and/or nonlinear unknown potentials.

From a more rigorous mathematical perspective,
it will be interesting to attempt to connect the present
setting to the extensive developments in treating integrable problems
with suitable boundary conditions (e.g. on the half line), as detailed
e.g., in Ref.~\cite{fokas}. A way to make this connection may be to consider
the GP equation e.g. solely on the attractive domain with a boundary
condition inferred by the incidence of the Gaussian wavepacket 
at $x=0$ (i.e., a Gaussian in time boundary condition). A potential
by-product of such a formulation might be the identification of the
number of solitary waves that will emerge, as a function of the
properties of this effective (and localized in time) boundary drive.

Finally, it would be of particular interest to extend considerations
to the two- or higher-dimensional setting. There, understanding the 
properties of the formed solitons, e.g., on the attractive interaction 
``domain'', taking into consideration 
the collapse feature that arises in the critical or super-critical
higher-dimensional case~\cite{sulem}, would be especially relevant.

\section*{Acknowledgements.} The work of D.J.F. was partially supported by the 
Special Account for Research Grants of the University of Athens.
P.G.K. acknowledges support from the National Science Foundation
under grant DMS-1312856, from FP7-People under grant
IRSES-605096 from
the Binational
(US-Israel) Science Foundation through grant 2010239,
and from the US-AFOSR under grant FA9550-12-10332.  
P.K. acknowledges support by EPSRC 
(grant EP/I017828/1) and the EU. P.G.K. and P.K. also acknowledge
the hospitality of the Synthetic Quantum Systems group 
and of Markus Oberthaler
at the Kirchhoff Institute for Physics (KIP) at the University
of Heidelberg, as well as that of the Center for Optical 
Quantum Technologies (ZOQ) and of Peter Schmelcher at
the University of Hamburg, as well as of the Center for Nonlinear
Studies at the Los Alamos National Laboratory.
P.G.K.'s work at Los Alamos is supported in part by the
U.S. Department of Energy.

\end{document}